\documentclass[11pt,a4paper]{aa}
\usepackage[utf8]{inputenc}
\usepackage[T1]{fontenc}
\usepackage{lmodern}
\usepackage{microtype}
\usepackage{setspace}
\usepackage{graphicx}
\usepackage{booktabs}
\usepackage{hyperref}
\hypersetup{colorlinks=true,linkcolor=blue,citecolor=blue,urlcolor=blue}
\usepackage{titling}
\usepackage{fancyhdr}
\usepackage{amsmath,amssymb}
\usepackage{afterpage}
\usepackage{multicol}
\usepackage{xcolor}
\usepackage{bibentry}

\begin{document}

\begin{titlepage}
  \centering
  \vspace*{0cm}
  {\Huge\bfseries Expanding Horizons \\[6pt] \Large Transforming Astronomy in the 2040s \par}
  \vspace{0.5cm}

  {\LARGE \textbf{White Dwarf Binaries: Probes of Future Astrophysics}\par}
  \vspace{0.3cm}

  \begin{tabular}{p{4cm}p{13.6cm}}
    \textbf{Scientific Categories:} & (Stars: binaries, evolution; white dwarfs; time-domain astronomy) \\
    \\
    \textbf{Submitting Author:} & Anna F. Pala  \\
    & European Southern Observatory, Karl Schwarzschild Straße 2, D-85748, Garching, Germany \\
    & annafrancesca.pala@eso.org\\
    \\
    \textbf{Contributing authors:} & 
Roberto Raddi$^{1}$,
Alberto Rebassa-Mansergas$^{1,2}$,  
Boris T. G{\"a}nsicke$^{3}$,
Richard I. Anderson$^{4}$,
Diogo Belloni$^{5,6}$, 
Avraham Binnenfeld$^{4}$,
Elm{\'e} Breedt$^{7}$, 
David Buckley$^{8}$,
Tim Cunningham$^{9}$,  
Alessandro Ederoclite$^{10}$,
Ana Escorza$^{11,12}$,
Valeriya Korol$^{13}$,  
Thomas Kupfer$^{14,15}$,
Domitilla de Martino$^{16}$, 
Jaroslav Merc$^{11,17}$,
Joaquin Meza$^{5,18}$,
Steven Parsons$^{19}$, 
Ingrid Pelisoli$^{3}$, 
Nicole Reindl$^{20}$, 
Pablo Rodr\'iguez-Gil$^{11,12}$, 
Alejandro Santos-Garc\'ia$^{1}$,
Simone Scaringi$^{21}$,
Paula Szkody$^{22}$, 
Odette Toloza$^{5,18}$,
Santiago Torres$^{1,2}$,  
Murat Uzundag$^{23}$, 
Monica Zorotovic$^{24}$
\vspace{0.13cm}\\
\multicolumn{2}{l}{\scriptsize $^{1}$Departament de Física, Universitat Politècnica de Catalunya, c/Esteve Terrades 5, 08860, Castelldefels, Spain}\\
\multicolumn{2}{l}{\scriptsize $^{2}$Institut d'Estudis Espacials de Catalunya (IEEC), C/Esteve Terradas, 1, Edifici RDIT, 08860, Castelldefels, Spain}\\
\multicolumn{2}{l}{\scriptsize $^{3}$Department of Physics, University of Warwick, Coventry, CV4 7AL, UK}\\
\multicolumn{2}{l}{\scriptsize $^{4}$Institute of Physics, \'Ecole Polytechnique F\'ed\'erale de Lausanne (EPFL), Observatoire de Sauverny, 1290 Versoix, Switzerland}\\
\multicolumn{2}{l}{\scriptsize $^{5}$Departamento de F\'isica, Universidad T\'ecnica Federico Santa Mar\'ia, Av. Espa\~na 1680, Valpara\'iso, Chile}\\
\multicolumn{2}{l}{\scriptsize $^{6}$S\~ao Paulo State University (UNESP), School of Engineering and Sciences, Guaratinguet\'a, Brazil}\\
\multicolumn{2}{l}{\scriptsize $^{7}$Institute of Astronomy, University of Cambridge, Madingley Road, Cambridge CB3 0HA, UK}\\
\multicolumn{2}{l}{\scriptsize $^{8}$South African Astronomical Observatory, Cape Town, 7935, South Africa}\\
\multicolumn{2}{l}{\scriptsize $^{9}$Center for Astrophysics, Harvard \& Smithsonian, 60 Garden St., Cambridge, MA 02138, USA} \\
\multicolumn{2}{l}{\scriptsize $^{10}$Centro de Estudios de F\'isica del Cosmos de Arag\'on (CEFCA), Plaza San Juan 1, 44001, Teruel, Spain}\\
\multicolumn{2}{l}{\scriptsize $^{11}$Instituto de Astrofísica de Canarias, E-38205 La Laguna, Tenerife, Spain}\\
\multicolumn{2}{l}{\scriptsize $^{12}$Departamento de Astrofísica, Universidad de La Laguna, E-38206 La Laguna, Tenerife, Spain}\\
\multicolumn{2}{l}{\scriptsize $^{13}$Max-Planck-Institut f\"ur Astrophysik, Karl-Schwarzschild-Stra{\ss}e 1, 85748, Garching, Germany}\\
\multicolumn{2}{l}{\scriptsize $^{14}$Department of Physics and Astronomy, Texas Tech University, 2500 Broadway, Lubbock, TX 79409, USA}\\
\multicolumn{2}{l}{\scriptsize $^{15}$Hamburg Observatory, University of Hamburg, Gojenbergsweg 112, 21029 Hamburg, Germany}\\
\multicolumn{2}{l}{\scriptsize $^{16}$INAF - Osservatorio Astronomico di Capodimonte, Salita Moiariello 16, I-80131 Naples, Italy}\\
\multicolumn{2}{l}{\scriptsize $^{17}$Astronomical Institute of Charles University, V Hole\v{s}ovi\v{c}k{\'a}ch 2, 180 00 Prague, Czech Republic}\\
\multicolumn{2}{l}{\scriptsize $^{18}$Millennium Nucleus for Planet Formation, NPF, Valparaíso, 2340000, Chile}\\ 
\multicolumn{2}{l}{\scriptsize $^{19}$Astrophysics Research Cluster, School of Mathematical and Physical Sciences, University of Sheffield, Sheffield S3 7RH, UK}\\
\multicolumn{2}{l}{\scriptsize $^{20}$Landessternwarte Heidelberg, Zentrum für Astronomie, Ruprecht-Karls-Universität, Königstuhl 12, 69117 Heidelberg, Germany}\\
\multicolumn{2}{l}{\scriptsize $^{21}$Department of Physics, Centre for Extragalactic Astronomy, Durham University, South Road, Durham DH1 3LE, UK}\\
\multicolumn{2}{l}{\scriptsize $^{22}$Department of Astronomy, University of Washington, Seattle, WA 98195-1700, USA}\\
\multicolumn{2}{l}{\scriptsize $^{23}$Institute of Astronomy, KU Leuven, Celestijnenlaan 200D, Leuven, 3001, Belgium}\\
\multicolumn{2}{l}{\scriptsize $^{24}$Instituto de F\'isica y Astronom\'ia, Universidad de Valpara\'iso, Gran Bretaña 1111, Playa Ancha, Valparaíso, 2360102, Chile}\\[0.1cm]\\
  \begin{minipage}{0.96\textwidth}
  \begin{center} \textbf{ABSTRACT} \end{center}
  White dwarf binaries are fundamental astrophysical probes. They represent ideal laboratories to test the models of binary evolution, which also apply to the sources of gravitational waves, whose detection led to the award of the 2017 Nobel Prize in Physics. Moreover, their final fate is intimately linked to Type Ia Supernovae (SNe\,Ia), i.e. the thermonuclear explosion of a white dwarf following the interaction with a companion star, which have become the fundamental yardsticks on cosmological distance scales and led to the discovery of dark energy and the award of the 2011 Nobel Prize in Physics. Finally, white dwarf binaries play a crucial role in influencing star formation and chemical evolution of the Galaxy by injecting energy into, and enriching, the interstellar medium with material ejected during nova eruptions and SN\,Ia explosions.
In the next decade, the advent of the Large Synoptic Survey Telescope (LSST) at the Vera Rubin Observatory will lead to the discovery of hundreds of thousands of white dwarf binaries. Nonetheless, the intrinsic faintness of the majority of these systems will prevent their spectroscopic characterisation with the instruments available in the 2030s. Hence ESO's Expanding Horizons call is timely for planning a future transformative facility, capable of delivering phase-resolved spectroscopic observations of faint white dwarf binaries, which are key to advancing our understanding of stellar and Galactic evolution and cosmology.

  \end{minipage}
  \end{tabular}
\end{titlepage}

\section{Introduction and Background}
\label{sec:intro}
White dwarfs represent the most common endpoint of stellar evolution. About 20\% of the currently observed white dwarfs reside in binaries \citep[either detached or accreting,][]{Torres+2022} and are fundamental to our understanding of a wide range of key astrophysical phenomena.

Of particular importance are white dwarf binaries with orbital periods from a few minutes to several days. They descend from main-sequence binaries that undergo at least one common envelope phase and their evolution (as that of all types of binaries) is driven by orbital angular momentum losses. The binary interaction often leads to the release of both kinetic energy and (potentially chemically enriched) material in the surrounding interstellar medium (e.g. during nova eruptions or Type\,Ia Supernova explosions, \citealt{Gehrz+1998,Matteucci+2021}). Therefore, white dwarf binaries influence star formation, Galactic chemical evolution, and they are also intimately linked to Type Ia Supernovae (SNe\,Ia), which are crucial standard candles for measuring distances on cosmological scales \citep{Branch+1992}. Finally, close double white dwarf binaries are among the brightest sources of low-frequency gravitational waves \citep{Nelemans2009,Scaringi+2023,Kupfer+2024}, making them vital targets for future space-based detectors, such as the \emph{Laser Interferometer Space Antenna} (\emph{LISA}, \citealt{Amaro-Seoane+2023}), the \emph{Tian-Quin} mission \citep{Luo+2016}, and the \emph{Lunar Gravitational-wave Antenna} (\emph{LGWA}, \citealt{Ajith+2025}).

\subsection{The population of white dwarf binaries}
\begin{figure*}
    \centering
    \vspace*{-12mm}
    \includegraphics[width=\textwidth]{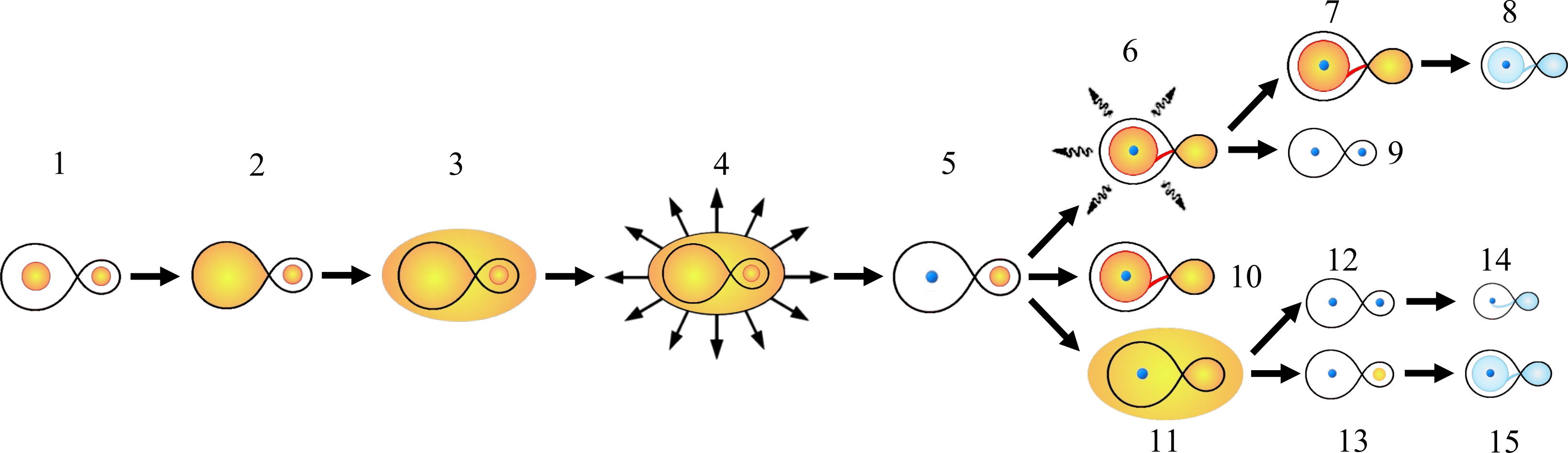}
    \caption{Evolution of white dwarf binaries: in a main-sequence star binary (1), the more massive star evolves first and fills its Roche lobe (2), leading to the formation of a common envelope (3). The envelope ejection (4) leaves behind a PCEB (5). If the companion is more massive than the white dwarf, it fills its Roche lobe when it leaves the main sequence. The systems undergo a phase of high rate mass transfer, during which the binary is observed as a super-soft X-ray source (6), leading either to a CV with a nuclearly evolved donor (7), or a double white dwarf (9). The CV later evolves in an AM\,CVn (8). If the companion is less massive, angular-momentum losses shrink the orbit, bringing it in contact with its Roche lobe. The system can then become a CV (10), or enter a second common envelope phase (11). If, at its onset, the companion core is degenerate, a double white dwarf is formed (12), otherwise, a helium star–white dwarf binary is formed (13). Subsequent Roche-lobe overflow produces an AM\,CVn with either a white dwarf (14) or helium star (15) donor.}
    \label{fig:wdbs}
\end{figure*}

White dwarf binaries descend from main-sequence binaries (Figure~\ref{fig:wdbs}) in which orbital angular momentum losses cause the orbit to shrink, possibly leading to a mutual interaction between the two stellar components. For sufficiently tight orbits ($a\,\lesssim\,10\,$au, \citealt{Farihi+2010}), the system undergoes a common envelope phase once the more massive star evolves off the main sequence. Friction extracts angular momentum from the orbit \citep[e.g.][]{Ivanova+2013,Ropke+2023}, leaving behind white dwarf plus main sequence star binaries with orbital periods ($P_\mathrm{orb}$) of hours to days \citep{Willems+2004,Santos-Garcia+2025}. In these post-common envelope binaries (PCEBs), companion stars can be M-dwarfs, brown dwarfs, or stars of spectral type A, F, G or K (the so-called WD+AFGK binaries).

In PCEBs, the orbital separation continues to decrease, finally causing the main sequence star to fill its Roche lobe and transfer mass onto the white dwarf via the inner Lagrangian point, typically with the formation of an accretion disc. These systems are known as accreting white dwarfs (AWDs) and come in different flavours \citep[e.g.][]{Belloni+2023}. For $P_\mathrm{orb} \gtrsim 60\,$min, AWDs are also known as cataclysmic variables (CVs) and have main-sequence (mostly late M-dwarfs) and brown-dwarf donors. In a small fraction ($\simeq 5\%$, \citealt{Pala+2020}) of CVs, the donor is initially more massive than the white dwarf (such as in WD+AFGK binaries). In these systems, unstable mass transfer at very high rates ensues, resulting in CVs with nuclear evolved (N/C enriched) companions of K and G spectral types \citep{Schenker+2002}. AWDs with degenerate donors also exist (the so-called AM\,CVns) and populate the shortest orbital periods ($P_\mathrm{orb} < 60\,$ min), hosting either helium stars or helium-core white dwarfs as donors.

Finally, double white dwarfs, i.e. binaries composed of two non-interacting white dwarfs, are thought to form following a phase of high rate mass transfer in the evolution of WD+AFGK binaries \citep{Zorotovic+2014}, or after a second common envelope phase \citep{Iben1991,Marsh1995}.

\section{Key Science Drivers in the 2040s}
\subsection{White dwarf binaries as probes of binary evolution} 
Orbital angular momentum losses and the common envelope phase are key ingredients in modelling the evolution of all compact binaries. The development of such models is currently limited by the lack of (i) accurate observational constraints on the efficiency of the common envelope phase \citep{DiStefano+2023}, and (ii) a complete theoretical framework describing magnetic wind braking, one of the main mechanisms for angular momentum loss \citep{Barraza-Jorquera+2025}. Statistically significant and well-defined samples of white dwarf binaries are one of the most powerful tools to constrain and further develop the models, as they allow carrying out a direct comparison between the observations and the prediction of binary population synthesis studies \citep{Pala+2020,Inight+2021,Rodriguez+2025,Torres+2025}.

\subsection{White dwarf binaries as gravitational wave sources} 
With an estimated population of $\simeq 100$ millions \citep{Marsh2011}, double white dwarfs are the dominant source of low-frequency gravitational waves in our Galaxy, particularly in the millihertz range where \emph{LISA}, \emph{Tian-Quin}, and \emph{LGWA} will be sensitive. While thousands of these systems will be individually detected, many millions will be unresolved and will produce a ``Galactic gravitational wave foreground'' \citep{Nelemans2009}, which will limit the sensitivity of \emph{LISA}. This is also the case for AWDs, which contribute with an additional signature into confusion noise \citep{Scaringi+2023}. Both foreground signals need to be precisely modelled for the correct interpretation of the data delivered by future gravitational wave missions \citep{Korol+2022,Kupfer+2024}. 

\subsection{White dwarf binaries as SN\,Ia progenitors} 
SNe\,Ia are thermonuclear explosions taking place in binaries following the interaction of a white dwarf with its companion. Many models of SNe\,Ia have been proposed (see \citealt{Wang+2012,Maoz+2014} for reviews). In the single-degenerate scenario, a white dwarf surpasses the Chandrasekhar mass limit following accretion from a non-degenerate donor \citep[e.g.][]{Iben+1984}. Another possibility is the double-detonation scenario, in which the detonation of a thick helium layer accreted onto the carbon-oxygen primary star triggers the ignition, leading to the explosion even at sub-Chandrasekhar masses \citep{Taam1980}. Finally, in the double-degenerate scenario, two white dwarfs in a binary lose orbital energy via gravitational wave emission and spiral inward. If their combined mass exceeds the Chandrasekhar limit, depending on their core compositions, the merger might lead to a thermonuclear explosion \citep{Shen+2024}. The mass of the white dwarf plays a vital role in assessing the final evolution of these binaries and it is critically important to reveal the pathway to SN\,Ia explosions. However, the current census of accurate mass measurements is too limited \citep{Napiwotzki+2020,Munday+2025} to deliver any meaningful comparison against the prediction of population models, SN\,Ia rates and delay time distributions, and it still remains unclear which configurations can lead to a successful explosion and whether multiple pathways coexist \citep{Maoz+2014}. 

\subsection{White dwarf binaries as tracers of ISM enrichment} 
In AWDs, the layer of accreted material is unstable and, along with part of the underlying white dwarf core, it is periodically ejected in the surrounding space via thermonuclear runways (``classical nova eruptions''). The ejected material efficiently enriches the interstellar medium (ISM) with elements synthesised during the eruptions, such as CNO-cycle products, Li, Be, and also contributes to dust production \citep{Gehrz+1998,Izzo+2015,Izzo+2018}.
Similarly, SN\,Ia explosions are a primary source of iron and other heavy elements (Ni and Mn), and are also a significant channel of energy injection into the ISM \citep{Matteucci+2009}. Both phenomena influence star formation and the chemical evolution of galaxies. Understanding their mechanisms, rates and delay-time distributions, is crucial for determining our understanding of the formation histories of galaxies \citep{Matteucci+2021}.

\section{From the next decade to the 2040 landscape}
\label{sec:openquestions}
\subsection{White dwarf binaries in the next decade}
\begin{figure}
    \centering
    \includegraphics[width=\linewidth]{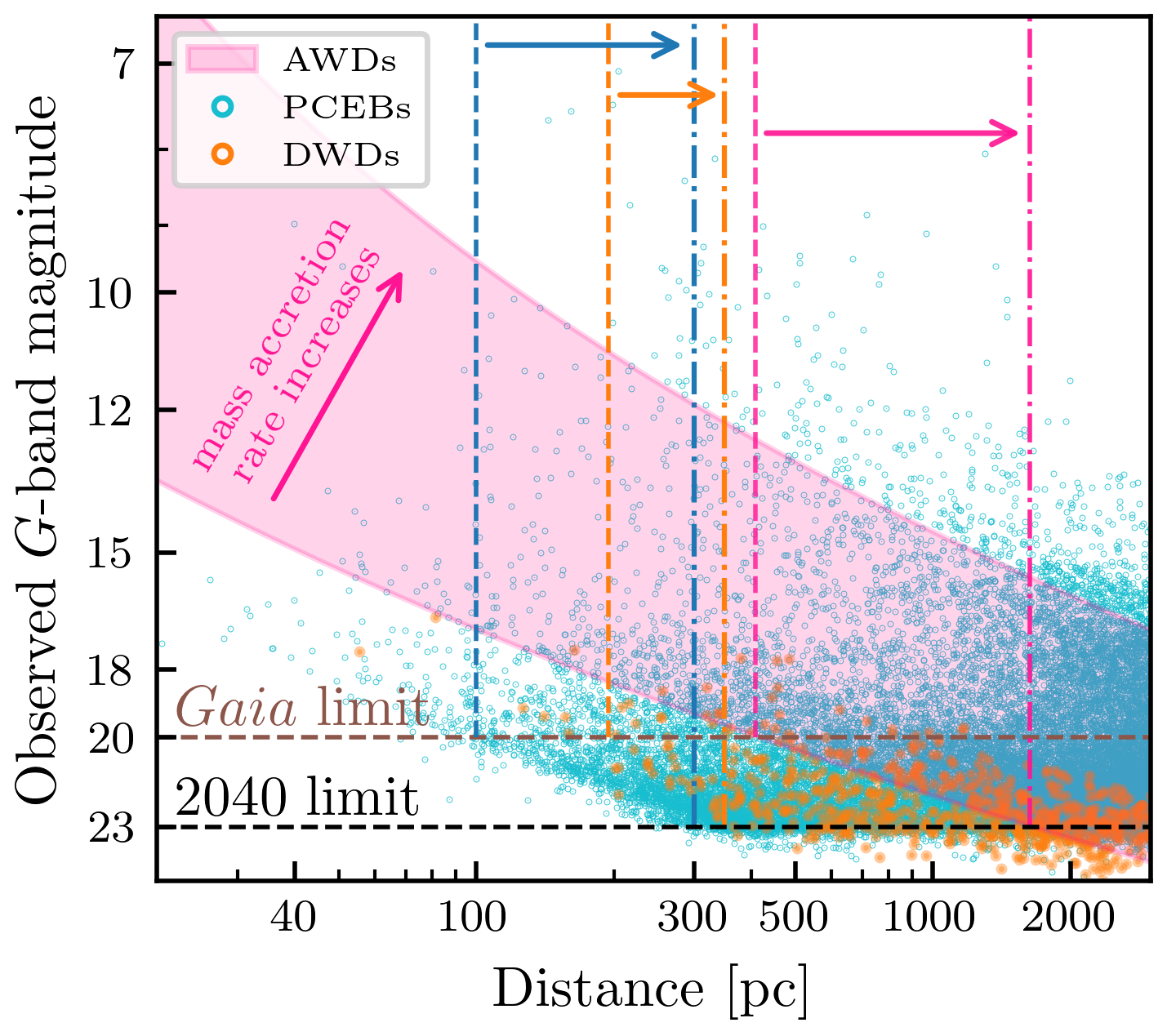}
    \caption{Synthetic magnitudes for PCEBs (cyan) and double white dwarfs binaries (orange) within 3\,kpc and brighter than $G \simeq 23$, computed using the Monte Carlo simulator from \cite{Santos-Garcia+2025}. AWDs are accretion-powered, and their typical magnitudes (pink area) depend on the accretion rate. \emph{Gaia} provides accurate magnitudes down to $G \simeq 20$ (dashed brown line), thus allowing obtaining volume-limited complete samples of PCEBs, double white dwarfs and AWDs out to $\simeq 100\,$pc (dashed blue line), $\simeq 150\,$pc (dashed orange line) and $\simeq 400\,$pc (dashed pink line), respectively. In the 2040s, we envisage a new facility capable of achieving $\mathit{SNR} \gtrsim 5$ down to $G \simeq 23$ (dashed black line), thus allowing to probe larger volumes, out to $\simeq 300\,$pc, $\simeq 350\,$pc and $\simeq 1600\,$pc, for PCEBs (dot-dashed blue line), double white dwarfs (dot-dashed orange line) and AWDs (dot-dashed pink line), respectively.}
    \label{fig:volumes}
\end{figure}

The upcoming \emph{Gaia} Data Releases 4 and 5 (December 2026 and late 2030, respectively) and the advent of wide-area multi-object spectroscopic instruments such as 4MOST \citep{deJong+2019}, WEAVE \citep{Jin+2024}, SDSS-V \citep{Kollmeier+2025} and DESI \citep{Cooper+2023}, along with deep, time-domain surveys like the Large Synoptic Survey Telescope (LSST, \citealt{Ivezic+2019}) will provide an unprecedented volume of astrometric, photometric, and spectroscopic data for hundreds of thousands of white dwarf binaries. However, these surveys will be limited to high-resolution spectroscopic only for sources with $G \simeq 16$, and low-resolution identification spectra down to $G \simeq 20$. As shown in Figure~\ref{fig:volumes}, this restricts the accessible white dwarf binary population to within $\simeq 100-300\,$pc. This sample will offer a first insight into the global properties of the population, yielding an estimate of space density and scale height, which enable indirect estimates of ages and formation rates and are expected to drive substantial progress in our general understanding of binary evolution. 

Nonetheless, although LSST will provide parallaxes for fainter targets ($G < 23$, \citealt{Fantin+2021}) and thereby the possibility to probe substantially larger volumes, these systems will remain inaccessible to the spectroscopic facilities available in the 2030s. This will prevent the characterisation of the individual sub-populations, which are dominated by intrinsically faint systems (e.g. CVs with brown dwarf donors, AM\,CVns and double white dwarfs), and, in turn, limit the development of detailed models of binary evolution.

Finally, next-decade facilities also lack the capability for phase-resolved observations optimised for the short orbital periods (5\,min - 1\,d) of the most compact white dwarf binaries. These are essential for measuring the binary parameters (such as orbital period and masses), investigating the final fates of white dwarf binaries and their connection to SNe\,Ia, as well as modelling the gravitational wave foreground. This is particularly constraining for double white dwarfs. Since their spectral energy distribution is often dominated by one of the two stellar components and their spectra resemble those of single white dwarfs, their characterisation requires high-resolution ($R>20\,000$) spectroscopic observations in order to resolve the narrow core of the H$\alpha$ line \cite[see e.g.][]{Napiwotzki+2003}. Achieving high resolution with UVES and CUBES \citep{Barbuy+2014}, restricts studies to only targets brighter than $G \simeq 16-18$ (depending on the orbital period). Looking further ahead, ANDES \citep{Maiolino+2013} at the ELT will push the limit to $G \simeq 18-20$, offering transformative capabilities for detailed follow-up studies of double white dwarfs. However, this will still restrict the sample to a small number of objects, since based on the number of white dwarfs observed by UVES over the past 20 years ($\simeq 1000$), we can expect ANDES to observe at best $\simeq 250$ by 2040, using its first five years of operation.

\subsection{Expanding the horizon of white dwarf binaries studies}
The key goal is to obtain spectroscopy of all compact white dwarf binaries ($G < 23$) identified by LSST and that will not be accessible to spectroscopic surveys available in the 2030s. Identification spectra and phase-resolved observations should be obtained not only for new discoveries but also for all systems previously identified by previous spectroscopic surveys for which phase-resolved data are lacking. This wealth of data will allow (i) establishing statistically significant volume-limited samples of the different sub-populations (PCEBS, AFGK+WDs, AWDs and double white dwarfs), as well as (ii) deriving orbital periods, masses, effective temperatures, abundances and rotation rates for individual systems. 

A dataset such as this will provide a comprehensive picture of white dwarf binary evolution at different stages and deliver stringent constraints on the evolutionary models. One of the key parameters is the space density and by comparing those of PCEBs, AWDs and double white dwarfs, it will be possible to infer the efficiency of the common envelope phase. 

Space densities, orbital periods and masses will allow an accurate constraint on the expected number of source detections in order to model the gravitational wave foreground that will limit the sensitivity of future gravitational wave missions. Accurate masses will also provide the ability to trace the white dwarf mass at different stages in the life of the binary and finally reveal the pathway to SN\,Ia explosions.

Finally, by exploring a larger volume (out to $\simeq 1.5\,$kpc), the scale height of each sub-population can be constrained, which is a proxy for the age of the binary. In this way, it will be possible to investigate the formation history of compact white dwarf binaries and its connection to different metallicity environments (halo vs. thin and thick disc), which directly affects the yields of processed material returned to the ISM by nova eruptions and SNe\,Ia.

\subsection{Technical requirements for the 2040s}
\begin{itemize}
\item A large multiplexing capability to observe thousands of objects efficiently. This can be achieved either with a survey telescope with a large field of view, or with a facility that combines smaller individual telescopes, which can be used together or individually on single targets (depending on their brightness).
\item The capability for high-cadence phase-resolved observations is crucial to measure orbital periods, radial velocities and the derivation of the binary parameters.
\item A large aperture ($> 8\,$m, either with a single mirror or with the combination of smaller telescopes), high sensitivity (especially at the shortest wavelength, down to the atmospheric cut-off at $\lambda \simeq 300\,$nm) and read-out noise-free detectors are required to achieve signal-to-noise ratios $\mathit{SNR} \gtrsim 5$ in short exposures ($\simeq 1-5\,$min). These are set by the temporal resolution necessary to avoid orbital smearing for the most compact binaries (exposure times of $\simeq 1/8$ of their $\simeq 5-10\,$min orbital periods).
\item One low ($R \simeq 5\,000$) and one high ($R > 20\,000$) resolution channel for AWDs and double white dwarfs, respectively.
\item The low-resolution channel should provide wide wavelength coverage, from the atmospheric cut-off in the blue ($\lambda \simeq 300\,$nm) to the $K$-band in the near-infrared ($\lambda \simeq 2500\,$nm). This is necessary to observe, in a single exposure, the different sources of emission in AWDs: the white dwarf (dominating the near-ultraviolet), the accretion disc (dominating the optical) and the companion star (dominating the near-infrared), as well as the two binary components in PCEBs. The high-resolution channel should cover the entire Balmer series.
\item An Integral Field Unit (IFU) should be available to probe the immediate environment around white dwarf binaries and to detect and characterise nebulosities that trace the ejecta of common envelopes, nova shells, and other structures, such as circumbinary discs and bow shocks, all of which (i) reflect possible mass-loss mechanisms that influence binary evolution, and (ii) trace how white dwarf binaries contribute to the ISM enrichment.
\end{itemize}

\begin{acknowledgements}
RR acknowledges support from Grant RYC2021-030837-I and MEC acknowledges grant RYC2021-032721-I, both funded by MCIN/AEI/ 10.13039/501100011033 and by ``European Union NextGeneration EU/PRTR''.\\
This research was partially supported by the AGAUR/Generalitat de Catalunya grant SGR-386/2021 and the Spanish MINECO grant, PID2023-148661NB-I00.\\
This work was partially supported by the MINECO grant PID2023-148661NB-I00 and by the AGAUR/Generalitat de Catalunya grant SGR-386/2021.\\
RIA is funded by the Swiss National Science Foundation through an Eccellenza Professorial Fellowship (award PCEFP2\_194638). \\
This project has received funding from the European Union Horizon Europe Research and Innovation Action under grant agreement no. 101183153-WST. Views and opinions expressed are however those of the author(s) only and do not necessarily reflect those of the European Union or the European Research Executive Agency (REA). Neither the European Union nor the REA can be held responsible for them. \\
JaM was supported by the Czech Science Foundation (GACR) project no. 24-10608O. \\
AE acknowledges the financial support from the Spanish Ministry of Science and Innovation and the European Union - NextGenerationEU through the Recovery and Resilience Facility project ICTS-MRR-2021-03-CEFCA.
\end{acknowledgements}

\bibliographystyle{aa_my}
\bibliography{references}

\end{document}